\documentclass[twocolumn,showpacs,preprintnumbers,superscriptaddress]{revtex4}

\usepackage{amsfonts,amssymb,amsmath,amsthm}

\usepackage{dcolumn}
\usepackage{bm}

\usepackage{epsfig}

\begin{document}

\title{Quantum computation via translation-invariant operations on a chain of qubits}

\author{Robert Raussendorf}
\affiliation{California Institute of Technology,\\
Institute for Quantum Information, Pasadena, CA 91125, USA}

\date{\today}

\begin{abstract}
 A scheme of universal quantum computation on a chain of qubits is
 described that does not require local control. All the required 
 operations, an Ising-type interaction and spatially uniform 
 simultaneous one-qubit gates, are translation-invariant.   
\end{abstract}

\pacs{3.67.Lx, 3.67.-a}

\maketitle


\section{Introduction}
\label{intro}

Symmetry reduces complexity. In physical systems realizing quantum computers, the highest degree of symmetry is therefore not the most desirable. A quantum computer needs to be sufficiently simple and robust to be controllable in an experiment yet complex enough to be universal. 
One may therefore ask the question ``How much symmetry does a quantum computer allow for?''. In fact, a number of physical systems considered for the realization of a quantum computer such as optical lattices \cite{BA} or arrays of micro-lenses \cite{BE} are translation invariant, and the above question acquires a practical aspect. 

Quite surprisingly, it turns out that universal quantum computation can tolerate a fair amount of symmetry. Recently, a scheme of quantum computation using the rotation-invariant measurement of the total `spin' of two qubits as the only gate  has been devised \cite{Soyuz}. Furthermore, a translation-invariant computation scheme has been described \cite{Voll}. 

The computational power of translation-invariant or nearly translation
invariant quantum systems was revealed in Lloyd's \cite{SL} and
Watrous' \cite{WTR} work on quantum cellular automata (QCA). In
\cite{WTR} it was shown that a one-dimensional QCA can simulate any
quantum Turing machine. Translation invariance is broken only by the
initial state which encodes the program. The schemes \cite{SL} and
\cite{SB2} allow to simulate quantum circuits using a chain of qubits
with a generic translation-invariant interaction. They require
different species of qubits in a periodic arrangement and local
addressability at one end-point of the chain. In the scheme \cite{Nag}
such individual addressing is only required in the initialization. The method proposed in \cite{Voll} is completely translation invariant in space.  It requires homogeneous one- and two-local operations on  5-level systems. 

Here I describe a scheme for universal quantum computation via translation-invariant operations on a chain of qubits.  No individual addressability is required.  The scheme uses an Ising-type interaction and spatially uniform one-qubit gates. The qubits are all of the same species. Cold atoms in optical lattices \cite{BA}, where the requirement of local control adds to the overall technological challenge, are a candidate for the realization of the presented scheme.

\section{Constructive elements, universality and scalability}
\label{Construct}

Consider a one-dimensional chain of $N$ qubits initialized in the state $|00..0\rangle$ which is repetitively updated according to the transition function
\begin{equation}
	\label{transit}
  T=\left(\bigotimes_{i=1}^{N-1}\Lambda(Z)_{i,i+1}\right)
  \left(\bigotimes_{i=1}^N H_i \right).
\end{equation}
That is, in each elementary step of the evolution first a Hadamard
gate  is applied to each qubit and second, conditional
phase gates are simultaneously applied to all pairs of neighboring
qubits. This QCA transition function has previously been discussed  in
\cite{SW}. 

Between the transitions one may apply translation-invariant
unitary transformations of the form
\begin{equation}
  \label{pulses}
  U_A(\alpha) =
  \bigotimes_{i=1}^N\exp\left(i\frac{\alpha}{2}A_i\right),
\end{equation}
with $A \in \{X,Y,Z\}$. (Note that the subscript `$i$' labels the
site. The same operation is applied to each qubit.)
These requirements are equivalent to  bang-bang pulses of form (\ref{pulses}) and a permanent 
Ising-type interaction ${\cal{H}}=\sum_{i=1}^{N-1}|11\rangle_{i,i+1}\langle 11|$.

Let us first observe that
\begin{equation}
\label{rev}
  T^{N+1}=R,
\end{equation}
where $R$ is the reflection operator that sends the state of the qubit chain
into its mirror image. Thus, despite the fact that the qubits at the end
points need not be addressed, it is relevant that the chain {\em{has}}
ends. (For adaption of the scheme to a ring of qutrits, see
remark 1 in Section~\ref{CR}). A proof of (\ref{rev}) is given in the
appendix. 

Apart from its use in the computational scheme described here, the bit reversal operation $R$ is interesting in its own. Recently, proposals for both approximate and perfect bit reversal in qubit chains with a Heisenberg- and $XY$- interaction have been made; See \cite{Bose}-\cite{Stolze} and references therein. For perfect mirror reflection in an $XY$-chain the coupling strength needs to vary with position \cite{Alban}, but only mildly \cite{Stolze}.
 
Relation (\ref{rev}) represents a method to achieve spatial reflection in systems with an Ising-type interaction. Here, the interaction strength is independent of position, but additional stroboscopic pulses (\ref{pulses}) are required to realize the uniform Hadamard transformations. \medskip

\begin{figure}
	\begin{center}
	\epsfig{file=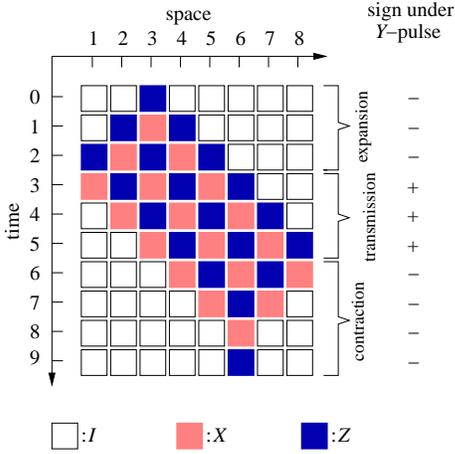, width=6cm}
	\caption{\label{evol}The evolution of a local Pauli
	observable, $Z_3$, for a qubit chain of  
	length $N=8$. The color-coded boxes denote Pauli
	 operators $X_i$ and $Z_i$, respectively,  and each row of
	boxes represents a tensor 
	 product of such Pauli operators. 
	Within the cycle that leads to reflection of the chain, each
	local observable  
	undergoes the phases of expansion, transmission and
	contraction. When expanding or   
	contracting, the operators pick up a sign factor of -1 under
	conjugation by a $Y$-pulse.}    
	\end{center}
\end{figure}

\paragraph{Universal set of gates.} The key to the construction of a universal set of gates for the described automaton is displayed in Fig.~\ref{evol}. The $N+1$-fold repetition of the elementary transition function constitutes half a clock cycle of the automaton. Within this period each local Pauli observable goes through the phases of expansion, transmission, and contraction to the mirror image of the initial position. During expansion and contraction, the propagated observable is susceptible to a global $Y$-pulse 
\begin{equation}
  Y:=U_Y(\pi)=\bigotimes_{i=1}^N Y_i.
\end{equation}
Namely, it picks up a sign factor under conjugation. Contrarily, during the phase of transmission where the observables behave as left- or right-movers \cite{SW} a $Y$-pulse has no effect. The duration of expansion, transmission and contraction phase depend on the initial position of the local Pauli observable. Therefore, each local Pauli observable shows a characteristic response to sequences of $Y$-pulses within the half-cycle. In this way, temporal control can be translated into spatial control. In the construction described below, suitably tailored sign-flips of Pauli observables are used to reverse rotation angles. Depending on whether a rotation angle is reversed or not, two matching rotations will either cancel or amplify each others effect. 

The degree of spatial control obtained suffices to simulate a quantum logic network. One constraint arises: every operation applied to the qubit chain is  reflection-symmetric. Therefore, qubit $i$ cannot be addressed separately from its mirror image at location $\overline{i}:=N+1-i$. To cope with this constraint, the same network is simulated twice on the chain, once on the left side and once---as mirror image---on the right. This doubles the required length of the chain and also influences the readout process, as will be discussed in Section~\ref{Construct}~$b$.

To probe the available spatial control,  consider the sequence
\begin{equation}
  \begin{array}{rcl}
    U_Z({\bf{c}},\alpha)&=& \left(T Y^{c_N} T
      Y^{c_{N-1}}T..TY^{c_1}TY^{c_0} U_Z(-\alpha/2) \right)\\
    && \left(T Y^{c_N} T
      Y^{c_{N-1}}T..TY^{c_1}TY^{c_0} U_Z(\alpha/2) \right),
  \end{array}
\end{equation}
where $\textbf{c}=(c_0, .. , c_N)^T$ is a binary vector. To analyze
the effect of this sequence, 
the Pauli operators $Y^{c_i}$ are propagated backwards in time until
they reach the $U_Z$-gates. There they accumulate to $\overline{Y}$,  
\begin{equation}
  \overline{Y}=(T^{-N}Y^{c_N}T^N)..(T^{-1}Y^{c_1}T)Y^{c_0}.
\end{equation}
Then, 
\begin{equation}
	\begin{array}{rcl}
	U_Z({\bf{c}},\alpha) &=& R \overline{Y}U_Z(-\alpha/2) R
	\overline{Y} U_Z(\alpha/2)\\ 
	&=&\bigotimes_{i=1}^N \exp\left(i s_i
	\frac{\alpha}{2}Z_i\right)	 
	\end{array}
\end{equation}
where we have used that $R\overline{Y}R=\overline{Y}$, and 
\begin{equation}
	\label{exchange}
	s_i=\left\{\begin{array}{ll}0, & \mbox{if}\;
    	[\overline{Y},Z_i]=0\\1,& \mbox{if}\;  \{\overline{Y},Z_i\}
    	=0\end{array}\right. .
\end{equation}
In this way, temporal control has been converted into spatial control, provided that---for suitable choices of ${\bf{c}}$---the binary variables $s_i({\bf{c}})$ do indeed vary with $i$.
	
The $s_i$ are easily computed in the
stabilizer formalism \cite{Go}-\cite{DeMoor}. Following \cite{DeMoor}, we write  Pauli operators $A$ in the form $i^\delta (-1)^\epsilon \tau_\textbf{a}$, where $\textbf{a}=\left(\begin{array}{c}\textbf{z}\\ \textbf{x}\end{array}\right)$ is a 2$N$-component binary vector, $\textbf{z}=(v_1,..,v_N)^T$, $\textbf{x}=(w_1,..,w_N)^T$;  $\delta, \epsilon \in \mathbb{Z}_2$, and $\tau_{\bf{a}}=\bigotimes_{i=1}^N
Z_i^{v_i} X_i^{w_i}$. The evolution of $A$ under conjugation by our Clifford unitary $T$, $A \longrightarrow T(A)=TAT^\dagger$, may then be followed in terms of $\delta$, $\epsilon$ and $\textbf{a}$. The scalars $\delta$ and $\epsilon$ have no influence on the sign factor $(-1)^{s_i}$ (\ref{exchange}), and we need to consider the update of $\textbf{a}$ only, $\textbf{a}(t) \longrightarrow \textbf{a}(t+1)= C\textbf{a}(t)$. Therein, $C$ is a $2N\times2N$ binary symplectic matrix which takes the form
\begin{equation}
  \label{C}
  C=\left(\begin{array}{cc}\Gamma& I\\I &0\end{array}\right).
\end{equation} 
$\Gamma$ is the adjacency matrix of the interaction graph (the line graph). Further, denote by $F$ the $2N\times 2N$-matrix $F=\left(\begin{array}{cc} 0& I\\I &0\end{array}\right)$ and observe that
$FC^{-1}=CF$. Now, the vector $\textbf{s}=(s_1,..,s_N)^T$ carrying the information about the sign flips under conjugation by $\overline{Y}$ is related to the vector $\textbf{c}$ describing the temporal sequence of $Y$-pulses, via $\textbf{s} = M_Z {\bf{c}}$. The matrix $M_Z$ encodes how temporal control is converted to spatial control. Its elements are given by
\begin{equation}
  \label{DefM_Z}
	\begin{array}{rcl}
  	M_Z(i,t)&=&\displaystyle{\textbf{a}_{Z_i}^T F C^{-t} \textbf{a}_Y}\\
	&=&\displaystyle{\textbf{a}_{Z_i}^T C^t\textbf{a}_Y.}
	\end{array}
\end{equation} 
The binary quantities $M_Z(i,t)$ can be defined for all $t$ ($i=1..N$). But the matrix $M_Z$ is 
the collection of $M_Z(i,t)$ only within the half-cycle $0\leq t \leq N$.

For the interval $t \in [-1,N]$ (including $[0,N]$) the $M_Z(i,t)$ take the values
\begin{equation}
  \label{spatial2}
  M_Z(i,t)=\theta(i-t-1)+\theta(t+i-N-1).
\end{equation}
Therein, the addition is modulo 2 and the step function $\theta$ is equal to 
one for all non-negative arguments, and zero otherwise. 

To prove (\ref{spatial2}) one may---in addition to $M_Z$---define a matrix $M_X$ that encodes the sign factor acquired by the $t$-steps-backward-propagated observable $X_i$ under conjugation with $\overline{Y}$, $M_X(i,t):=\textbf{a}_{X_i}^T  C^{t}\textbf{a}_Y$.  Then, because of $TX_iT^{-1}=Z_i$ and $TZ_iT^{-1}=X_i\bigotimes_{j|\,\Gamma_{ij}=1}Z_j$, the $M_Z(i,t)$ and $M_X(i,t)$ obey the recursion relations
\begin{equation}
  \label{Recrel}
  \begin{array}{rcl}
    M_Z(i,t+1) &=& \displaystyle{M_Z(i,t-1)+\sum_{j |\,\Gamma_{ij}=1}M_Z(j,t),}\\
    M_X(i,t+1) &=& M_Z(i,t).
  \end{array}
\end{equation}
The solution of the recursion relation for $M_Z(i,t)$ is unique once the boundary conditions on two consecutive time slices are specified. In the discussed case, the boundary conditions are $M_Z(i,-1)=M_Z(i,0)=1$, for all $i=1..N$. The expression on the r.h.s of (\ref{spatial2}) obeys the recursion relation (\ref{Recrel}) in the interval $0\leq t\leq N-1$ and the boundary conditions, and is thus the correct expression for $M_Z$. 

\begin{figure}[ht]
	\begin{center}
	\epsfig{file=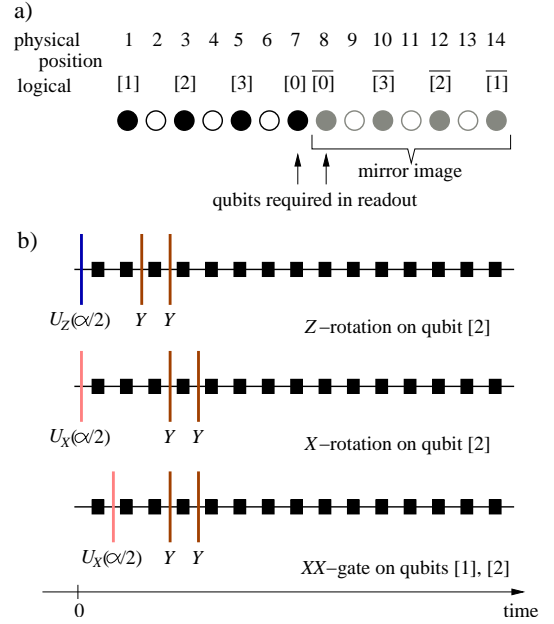,width=7cm}
	\caption{\label{Seq}Arrangement of logical qubits within the chain (a) and pulse 
	sequences for gates (b), for $n=3$. a) Only every second physical qubit in the
	 chain, ``$\bullet$'', is used as a 
	logical qubit. The qubits in between, ``$\circ$'',  are ancillas.
	b) Pulse sequences for the implementation of the universal gates, during the first half of 
	a clock cycle. The second half-cycle is analogous, with $\alpha \rightarrow -\alpha$.
	The symbol ``$\blacksquare$'' denotes application of the elementary transition 
	function $T$. The vertical bars denote pulses (\ref{pulses}).}
	\end{center}
\end{figure}

To perform a $z$-rotation on qubit $i$, one may apply a
$Y$-pulse at times $i-1$ and $i$. For $c_t=\delta(t,i-1)+\delta(t,i)$, with (\ref{spatial2}),
one obtains $s_j=\delta(j,i)+\delta(j,N+1-i)$. Therefore, the pulse for a $z$-rotation of qubit $i$ is   
\begin{equation}
  \label{Zgate}
  \begin{array}{rcl}
    e^{i\frac{\alpha}{2}Z_i}
    e^{i\frac{\alpha}{2}Z_{\overline{i}}}&=&
    \left(T^{N+1-i}YTYT^{i-1}U_Z(-\alpha/2)\right)\\&&
    \left(T^{N+1-i}YTYT^{i-1}U_Z(\alpha/2)\right),
  \end{array}
\end{equation}
with $\overline{i}=N+1-i$ the mirror site of $i$. In this way, two out of $N$ $z$-rotations can be selected. A further discrimination between two
sites $i$ and $\overline{i}$ is not possible because of the reflection
symmetry of every sequence $U({\bf{c}},\alpha)=R
U({\bf{c}},\alpha) R^{-1}$. As a result of this symmetry, each quantum algorithm is run on the qubit chain in two copies, one being the mirror image of the other. This accounts for a factor of two in spatial overhead.

From (\ref{Zgate}) the remaining universal gates can be
deduced easily. Relation (\ref{Zgate}) is conjugated by $T^{-1}$.  Using $T^{-1}Z_iT=X_i$, one 
finds 
\begin{equation}
  \label{Xgate}
  \begin{array}{rcl}
    e^{i\frac{\alpha}{2}X_i}
    e^{i\frac{\alpha}{2}X_{\overline{i}}}&=&
    \left(T^{N-i}YTYT^{i}U_X(-\alpha/2)\right)\\&&
    \left(T^{N-i}YTYT^{i}U_X(\alpha/2)\right).
  \end{array}
\end{equation} 
Conjugating relation (\ref{Xgate}) again by $T^{-1}$ and noting that $T^{-1}X_iT=Z_i\bigotimes_{j |\,\Gamma_{ij}=1}
  X_j=:K_i$ one obtains
\begin{equation}
  \label{XXgate}
  \begin{array}{rcl}
    e^{i\frac{\alpha}{2}K_i}
    e^{i\frac{\alpha}{2}K_{\overline{i}}}&=&
    \left(T^{N-1-i}YTYT^{i}U_X(-\alpha/2)T\right)\\&&
    \left(T^{N-1-i}YTYT^{i}U_X(\alpha/2)T\right).
  \end{array}
\end{equation}
Now, the length of the chain is doubled a second time. Specifically, the
logical qubits in state $|\psi\rangle$ are interlaced by ancilla
qubits which remain in the state $|0\rangle$ throughout the computation, 
\begin{equation}
  |\psi\rangle_{1,2,3,..,n} \longrightarrow
  |\psi^\prime\rangle= 
  |\psi\rangle_{1,3,5,..,2n-1}\otimes |0..0\rangle_{2,4,..,2n-2}.
\end{equation}
At this point it is suitable to introduce a logical coordinate $[j]$ that is related to the physical location within the qubit chain via $[j]=2j-1$, for $1 \leq j\leq n$, such that e.g. $|\psi^\prime\rangle=|\psi\rangle_{[1],[2],..,[n]}\otimes |0..0\rangle_{2,4,..,2n-2}$.  
Then, the action of the $K_{2j}$ gate is equivalent to a two-qubit next-neighbor entangling gate,
\begin{equation}
  \label{XX2}
  \exp\left(i\frac{\alpha}{2}K_{2j}\right)|\psi^\prime\rangle \equiv  
  \exp\left(i\frac{\alpha}{2}X_{[j]}X_{[j+1]}\right)|\psi^\prime\rangle.
\end{equation}
The gates (\ref{Zgate}), (\ref{Xgate}) and (\ref{XXgate}) form
a gate set that can be easily converted into the standard
universal set \cite{Boy}.\medskip

\paragraph{Readout.} As an example for global readout I consider a measurement of the $z$-component
 of the total spin, i.e., of the observable
\begin{equation}
	S_Z= \sum_i Z_i.
\end{equation}
This is a model for readout e.g. of atomic qubits via resonance fluorescence spectroscopy, if the atoms are well within the Lamb-Dicke limit (their separation and fluctuation of position are much smaller than the optical wave length).
In this case the underlying physics prohibits the extraction of local information. The readout method described below equally works for a scenario where a local readout could in principle be performed but is not pursued due to technological limitations. For specification of such a measurement model, see remark 3 in Section~\ref{CR}. 

With the capability to perform arbitrary unitary evolution on a quantum register the total spin observable $S_Z$ acting on that register may be conjugated into any desired observable, and the standard network readout of the individual $Z_i$ should be feasible. However, for the setting described here the readout procedure is, in the case of probabilistic algorithms, slightly complicated by the fact that two copies of the algorithm are run simultaneously and the readout measurement couples them. 

For the readout each of the two circuits on the chain requires one additional logical qubit that is in the
state $|0\rangle$ until the readout starts. The location of this qubit within the logical quantum register is denoted as $[0]$ which shall correspond to the physical position $2n+1$ in the qubit chain (see Fig.~\ref{Seq}a). The readout consists of three steps.
First, $S_Z$ is measured and the outcome $m$ is recorded. Second, for all logical qubits $j=1..n$ the following procedure is performed: 2.1) Apply CNOT gates $\Lambda(X)_{[j],[0]}\otimes \Lambda(X)_{\overline{[j]},\overline{[0]}}$ ([0], $\overline{[0]}$ are the target qubits). 2.2) Measure the observable $S_Z$ and record the outcome $m(j)$. 2.3) Apply the CNOT gates $\Lambda(X)_{[j],[0]}\otimes \Lambda(X)_{\overline{[j]},\overline{[0]}}$ again. Third, denote by $J=\{j_1, .. ,j_{|J|}\}$ those $j\leq n$ for which $m-m(j)=2$. If $|J|<2$ the readout is finished. Otherwise, for all $k=2..|J|$ perform the following procedure: 3.1) Apply the Toffoli-gates $\Lambda_2(X)_{[j_1],[j_k];[0]} \otimes \Lambda_2(X)_{\overline{[j_1]},\overline{[j_k]};\overline{[0]}}$ ([0], $\overline{[0]}$ are the target qubits), 3.2) Measure $S_Z$ and record the measurement outcome $m(j_1,j_k)$. 3.3)  Apply the Toffoli-gates $\Lambda_2(X)_{[j_1],[j_k];[0]} \otimes \Lambda_2(X)_{\overline{[j_1]},\overline{[j_k]};\overline{[0]}}$ again.

The conditional phase- and Toffoli gates are their own inverse such that the above protocol amounts to a sequence of measurements where the first one is of $S_Z$ and the following ones are of conjugated observables. These observables mutually commute because they are all diagonal in the computational basis, and the final state is a simultaneous eigenstate of them. The observables measured in step 2 are $Z_{[0]}Z_{[j]}+Z_{\overline{[0]}}Z_{\overline{[j]}}+\sum_{i \neq [0],\overline{[0]}}Z_i$. Since the qubits at locations $[0]$ and $\overline{[0]}$ are individually in the state $|0\rangle$, and the observable $S_Z=\sum_iZ_i$ already has been measured in step 1, step 2 effectively acts as the measurement of the observables $Z_{[j]}+Z_{\overline{[j]}}$. 

Similarly, step 3 amounts to the measurement of the observables $|11\rangle_{[j_1],[j_k]}\langle11| + |11\rangle_{\overline{[j_1]},\overline{[j_k]}}\langle11|$, $\forall k$. This measurement assigns a sharp value to $r_{[j_1]}r_{[j_k]} + \overline{r}_{[j_1]}\overline{r}_{[j_k]}$.  Its purpose is, in combination with the information gathered in step 2, to discriminate between the two cases $\{r_{[j_1]}=r_{[j_k]}, \overline{r}_{[j_1]}=\overline{r}_{[j_k]}\}$ and $\{r_{[j_1]} \neq r_{[j_k]}, \overline{r}_{[j_1]} \neq
\overline{r}_{[j_k]}\}$, for each $k$.

The state of the two quantum registers after the measurement, leaving out all the ancillas in the state $|0\rangle$, may be written as a superposition $\sum_\alpha c_\alpha |\textbf{r}^{(\alpha)}\rangle |\overline{\textbf{r}}^{(\alpha)}\rangle$ of computational basis states $|\textbf{r}^{(\alpha)}\rangle |\overline{\textbf{r}}^{(\alpha)}\rangle$, where $\textbf{r}=(r_{[1]},..,r_{[n]})^T$, $\overline{\textbf{r}}=(r_{\overline{[1]}}, .., r_{\overline{[n]}})^T$. The measurements then impose constraints upon which basis states are admissible, i.e., occur with nonzero coefficients $c_\alpha$. The constraints imposed by the measurements in step 2 read
\begin{equation}
	\label{constr1}
	\begin{array}{ll}
	r_{[j]}=\overline{r}_{[j]}=0,& \mbox{for}\,\, m-m(j)=0,\\
	r_{[j]}+\overline{r}_{[j]} \mod 2=1,& \mbox{for}\,\, m-m(j)=2,\\
	r_{[j]}=\overline{r}_{[j]}=1,& \mbox{for}\,\, m-m(j)=4,
	\end{array}
\end{equation}
for all $j=1..N$. The constraints that arise from the measurements in step 3 are 
\begin{equation}
	\label{constr2}
	\begin{array}{ll}
	r_{[j_1]}+r_{[j_k]} \mod 2=0,& \mbox{for}\,\, m-m(j_1,j_k)=2,\\
	r_{[j_1]}+r_{[j_k]} \mod 2=1,& \mbox{for}\,\, m-m(j_1,j_k)=0,
	\end{array}
\end{equation}
for $[j_1],[j_k] \in J$, $k=2..|J|$. The system (\ref{constr1}, \ref{constr2}) of linear equations has a unique solution if $J=\emptyset$ and two solutions otherwise. In the latter case, the two solutions $\textbf{r}^{(1)}$ and $\textbf{r}^{(2)}$ correspond to two states $|\textbf{r}^{(1)}\rangle|\textbf{r}^{(2)}\rangle$ and  $|\textbf{r}^{(2)}\rangle|\textbf{r}^{(1)}\rangle$, and the state of the quantum register after readout is their reflection-symmetric linear combination.  The solutions $\textbf{r}^{(1)}$ and $\textbf{r}^{(2)}$ are, modulo interchange, uniquely specified by the measurement outcomes $m$, $\{m(j)\}$ and $\{m(j_1,j_k)\}$.  For a probabilistic algorithm, the automaton produces two potential solutions in each run.\medskip

\paragraph{Overhead.} For the described method of universal 
computation via translation-invariant interaction it is relevant that the 
elementary transition function $T$ repeated a sufficient number of times reduces
to the identity operation. Such an elementary `clock cycle' takes
$2(N+1)$ time steps where $N$ is the length of the chain. But as
described so far, one still ends up with a set of local and
next-neighbor gates and thus pays the price of
next-neighbor slowdown twice. This is not necessary.  Before the resources are counted, a long-distance entangling gate is described. As compared to a logically equivalent composition of next-neighbor and local gates, it cuts the required operational resources by a factor of order $n$.

The natural gate set for the described scheme of quantum computation does not 
only contain local and next-neighbor gates, but all gates of the form
\begin{equation}
  \label{ElmGates}
  U_X(\alpha,L_1,L_2)=\exp\left(i\frac{\alpha}{2}
  \bigotimes_{l=L_1}^{L_2}X_{[l]}\right).
\end{equation}
 As before, the logical position $[l]$ corresponds to the physical location $2l-1$ in the chain. The sequence for implementing $U_X(\alpha,L_1,L_2)$ is
\begin{equation}
  \label{ElmX}
  \begin{array}{rcl}
    U_X(\alpha,L_1,L_2) &=& \left(T^{t_3}YTYT^{t_2}
    U_X(-\alpha/2)T^{t_1} \right)\\
    && \left(T^{t_3}YTYT^{t_2} U_X(\alpha/2)T^{t_1} \right),
  \end{array}
\end{equation}
with $t_1=L_2-L_1$, $t_2=L_1+L_2-2$ and $t_3=N-2L_1+2$.
Now note that, with $|-\rangle:=-X|-\rangle$, 
\begin{equation}
  \begin{array}{rcl}
  e^{i\pi|--\rangle_{l_1,l_2}\langle--|}&=&
  U_X(\pi/2,l_1,l_2) U_X(-\pi/2,l_1+1,l_2) \\
  &&  U_X(\pi/2,l_1+1,l_2-1)\\
  && U_X(-\pi/2,l_1,l_2-1).
  \end{array}
\end{equation}
Thus, a long-distance $x$-controlled spin-flip can be implemented with 4 
elementary gates (\ref{ElmX}). These four operations can be grouped into two pairs, with constant value of $L_2$ in each pair. Either pair of operations can be implemented in one clock cycle because the $Y$-pulse sequences of the two operations coincide. An $x$-controlled spin-flip thus takes two clock cycles irrespective of the distance between control and target qubit.\medskip

Now counting the resources, a quantum computation on $n$ qubits in the described scheme requires a qubit chain of length $N=4n+2$. Each clock cycle takes $8n+6$ elementary time steps (applications of $T$), and each of the universal gates $\exp\left(i\alpha/2\,Z_l\right)$, $\exp\left(i\alpha/2\,X_l\right)$ and $ e^{i\pi|--\rangle_{l_1,l_2}\langle--|}$ takes at most two clock cycles. As compared to the network model, there arises a constant overhead of $4$ in the spatial and a linear overhead of $16n$ in temporal resources.   

\section{Concluding remarks}
\label{CR}

Universal quantum computation can be performed by translation-invariant operations on a chain of qubits initialized in the state $|00..0\rangle$. The described method requires translation-invariant interaction of Ising type and spatially uniform local rotations. As compared to the network model, there occurs a constant spatial and a linear temporal overhead.

Three remarks: 1) In the described setting translation invariance is broken by the finite extension of the qubit chain. To create a setting with perfect translational invariance, one may replace the qubits by qutrits, where the additional level $|2\rangle$ does not take part in any interaction, and consider a $N+1$-qubit ring geometry instead of a chain. Set $|\varphi(i)\rangle=|0..0\rangle_{1..i-1}|2\rangle_i|0..0\rangle_{i+1..N+1}$. With either the translation-invariant superposition $\frac{1}{\sqrt{N+1}}\sum_{i=1}^{N+1}|\varphi(i)\rangle$ or mixture $\frac{1}{N+1}\sum_{i=1}^{N+1} |\varphi(i)\rangle \langle\varphi(i)|$ as the initial state the scheme works as before. \\
2) It may occur that a qubit chain capable of the required interaction has been created but its length $N$ is unknown. It can be found out easily by repetitions of the following protocol: Initialize the chain in the state $|00..0\rangle$, apply the transition function $T$ t times and subsequently measure the spin observable $S_Z=\sum_{i=1}^NZ_i$.  The received signal $\langle S_Z(t)\rangle$ carries a characteristic signature of $N$, namely $\langle S_Z(t)\rangle = N\delta(t \mod N+1)$.\\
3) Deviating from the setting discussed in Section~\ref{Construct}$\,b$, the available measurement may be such that local information could in principle be retrieved but is not retrieved, due to technological limitations. As an example, consider cold atoms in an optical lattice which are read out via resonance fluorescence spectroscopy and whose separation is larger than the wave length of the probing laser. In this case, the readout can be modeled by the map 
\begin{equation}
\rho\longrightarrow \rho^\prime(m)\sim \!\!\! \sum_{\{m_i\}|\,\sum_i m_i =m} P(\{m_i\}) \rho P(\{m_i\}),
\end{equation}
with the projectors 
$ P(\{m_i\})=\bigotimes_{i=1}^N\frac{1+m_i Z_i}{2}$, and $m_i=\pm 1\, \forall\, i=1..N$. Every atom is individually measured but only the {\em{global}} signal $m=\sum_i m_i$ is recorded. For measurements of this type the readout procedure is the same as discussed in Section~\ref{Construct} and has the same efficiency.

\begin{acknowledgments}

I thank Frank Verstraete, Luming Duan, GM Jeremia, Robin Blume-Kohout, Gavin Brennen, Ren\'e Stock and Sergey Bravyi for discussions. This work was supported by the Multidisciplinary University Research Initiative program under Grant No DAAD19-00-1-0374.
\end{acknowledgments}

\appendix
\section{Bit reversal}
\label{P1D}

Sufficiently many applications of the elementary transition function $T$ reverse the order of qubits within the chain. As stated in (\ref{rev}), with $R$ the spatial reflection, 
\begin{eqnarray}
	T^{N+1}=R. \nonumber
\end{eqnarray}
This phenomenon is related to the transmission of one-qubit states through chains described in \cite{Brenn}. Relation (\ref{rev}) is now proved. What needs to be shown is $T^{N+1}(X_p)=X_{\overline{p}}$ and  $T^{N+1}(Z_p)=Z_{\overline{p}}$, for all $p=1,\, ..\, ,N$. For this purpose, the vector space formulation \cite{Go}-\cite{DeMoor} of the stabilizer formalism is used. In particular, I use the conventions and results of \cite{DeMoor}. The evolution of the Pauli observables $X_p$, $Z_p$ is followed in terms of the binary quantities $\textbf{a}$ and $\epsilon$, introduced in Section~\ref{Construct} below Eq. (\ref{exchange}). 

\begin{figure}[thb]
	\begin{center}
	\epsfig{file=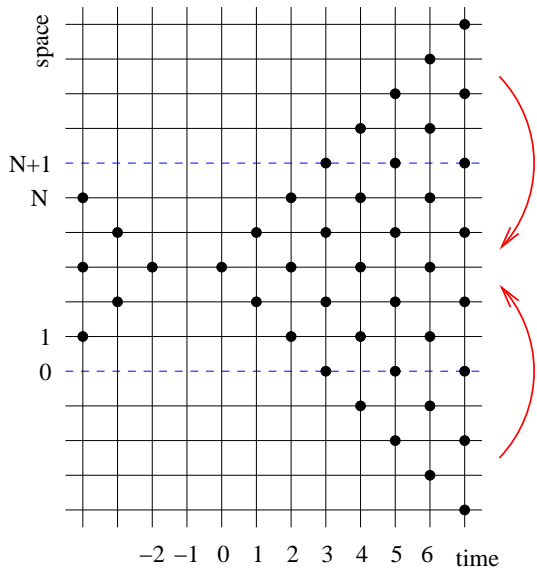,width=7cm}
	\caption{\label{foldback}The solution for the recursion 
	relations (\ref{infi}) in free space. Each vertex in the  lattice denotes a space-time 
	point $(i,t)$. The symbol ``$\bullet$" indicates $v_i(t)=1$ and ``$+$" $v_i(t)=0$. 
	To obtain the solution for the case with boundary from the free space solution, the 
	lattice is `folded back' to the space interval $[0,N+1]$. }	
	\end{center}
\end{figure}

First, the recursion relation $\textbf{a}(t+1)=C\textbf{a}(t)$ is translated into a recursion relation for the $\textbf{z}$-part of $\textbf{a}$ alone and a relation that expresses $\textbf{x}$ in terms of $\textbf{z}$.
From (\ref{C}), taking into account that $\Gamma$ represents a line graph, the following recursion relations are obtained for $\textbf{z}=(v_1,..,v_N)^T$ and $\textbf{x}=(w_1,..,w_N)^T$:
\begin{equation}
	\label{Recur}
	\begin{array}{rclr}
	v_1(t+1)&=&v_2(t)+v_1(t-1),&\\
	v_i(t+1)&=&v_{i-1}(t)+v_{i+1}(t)+v_i(t-1),& 1 <i< N,\\
	v_N(t+1)&=&v_{N-1}(t)+v_N(t-1),\\ \\
	w_i(t+1)&=&v_i(t),& 1\leq i \leq N.
	\end{array}
\end{equation}
We seek the solution of these recursion relations for the time $t=N+1$, with boundary 
conditions 
\begin{equation}
	\label{bc}
	v_i(-1)=0,\, v_i(0)=\delta(p-i), \, \forall\, 1\leq i \leq N.
\end{equation}
The translation-invariance of the recursion relations (\ref{Recur}) is broken by the finite extension of the chain. The problem of solving these recursion relations is now reduced to the translation-invariant case of the infinite chain. Note that if a configuration $\{\tilde{v}_i(t)\}$ obeys the recursion relations 
\begin{equation}
	\label{infi}
	\tilde{v}_i(t+1)=\tilde{v}_{i-1}(t)+ \tilde{v}_{i+1}(t)+ \tilde{v}_{i}(t-1),	
\end{equation}
then 
\begin{equation}
	\label{reduc}
	v_i(t):=\sum_{l=-\infty}^\infty \tilde{v}_{i+2l(N+1)}+\tilde{v}_{-i+2l(N+1)}
\end{equation}
obeys the recursion relation (\ref{Recur}). The reduction of the case with spatial boundary to the case without boundary is illustrated in Fig.~\ref{foldback}.

It is easily checked that
\begin{equation}
	\label{infsol}
	\tilde{v}_i(t)=\left(\theta(p-i+t)+\theta(r-i-t-1)\right)\delta(p-i+t \mod 2)
\end{equation}
is the solution of (\ref{infi}) with the boundary conditions 
$\tilde{v}_i(-1)=0,\, \tilde{v}_i(0)=\delta(p-i)$, $\forall i$. 
If this expression is inserted into the r.h.s. of (\ref{reduc}) 
one obtains a solution for $v_i(t)$ with boundary conditions (\ref{bc}). It is observed that
\begin{equation}
	\label{checkerboard}
	\begin{array}{rcl}
	v_i(t) &\sim& \delta(p-i+t \mod 2),\\
	w_i(t) &\sim& \delta(p-i+t+1 \mod 2).
	\end{array}
\end{equation}

Next, the solution  $v(t)$ (\ref{reduc}), (\ref{infsol}) is evaluated for $t=N+1$. One obtains
\begin{equation}
	v_i(N+1)=\delta(p-(N+1-i)).
\end{equation}
This implies $T^{N+1}(Z_p)=\pm Z_{\overline{p}}$. 

Next, the sign factor $(-1)^{\epsilon}$ in the above relation is worked out. Using the result of \cite{DeMoor}, the recursion relation for $\epsilon$ reads $\epsilon(t+1)=\epsilon(t)+\textbf{z}(t)^T \Gamma_L \textbf{z}(t)+\textbf{x}(t)^T\textbf{z}(t)$, with $\Gamma_L$ the lower triangular part of $\Gamma$. With (\ref{checkerboard}), $\epsilon(t)\equiv \epsilon(0)=0$ such that $T^{N+1}(Z_p)=Z_{\overline{p}}$. Finally, with $T(X_p)=Z_p$, $T^{N+1}(X_p)=T^{-1} T^{N+1} T (X_p)= X_{\overline{p}}$.


\end{document}